# Disentanglement of intrinsic and extrinsic side-jump scattering induced spin Hall effect in N-implanted Pt

Utkarsh Shashank[1], Yoji Nakamura[1], Yu Kusaba[1], Takafumi Tomoda[1], Razia Nongjai[2], Asokan Kandasami[2,3], Rohit Medwal[4], Rajdeep Singh Rawat[5], Hironori Asada[6], Surbhi Gupta[5], and Yasuhiro Fukuma[1,7*]

[1]*Department of Physics and Information Technology, Faculty of Computer Science and Systems Engineering, Kyushu Institute of Technology, 680-4 Kawazu, Iizuka 820-8502, Japan*

[2]*Materials Science Division, Inter University Accelerator Centre, Aruna Asaf Ali Marg, New Delhi 110067, India*

[3]*Department of Physics and Centre for Interdisciplinary Research, University of Petroleum and Energy Studies (UPES) Dehradun, Uttarakhand 248007, India*

[4]*Department of Physics, Indian Institute of Technology Kanpur, Kanpur 208016, India*

[5]*Natural Sciences and Science Education, National Institute of Education, Nanyang Technological University 637616, Singapore*

[6]*Graduate School of Sciences and Technology for Innovation, Yamaguchi University, 2-16-1 Tokiwadai, Ube 755-8611, Japan*

[7]*Research Center for Neuromorphic AI hardware, Kyushu Institute of Technology, Kitakyushu 808-0196, Japan*

[*]Correspondence should be addressed to Y.F. : fukuma@phys.kyutech.ac.jp

## ABSTRACT

The rapidly evolving utilization of spin Hall effect (SHE) arising from spin-orbit coupling in 5d transition metals and alloys have made giant strides in the development of designing low-power, robust and non-volatile magnetic memory. Recent studies, on incorporating non-metallic lighter elements such as oxygen, nitrogen and sulfur into 5d transition metals, have shown an enhancement in damping-like torque efficiency $\theta_{DL}$ due to the modified SHE, but the mechanism behind this enhancement is not clear. In this paper, we study $\theta_{DL}$ at different temperatures (100-293 K) to disentangle the intrinsic and extrinsic side-jump scattering induced spin Hall effect in N-implanted Pt. We observe a crossover of intrinsic to extrinsic side-jump mechanism as the implantation dose increases from $2\times10^{16}$ ions/cm$^2$ to $1\times10^{17}$ ions/cm$^2$. A sudden decrease in the intrinsic spin Hall conductivity is counterbalanced by the increase in the side-jump induced SHE efficiency. These results conclude that studying $\theta_{DL}$ as a function of implantation dose, and also as a function of temperature, is important to understand the physical mechanism contributing to SHE, which has so far been unexplored in incorporating non-metallic element in 5d transition metals.

## I. INTRODUCTION

Spin Hall effect (SHE) [1-4] has been garnering much attention in the development of spin-orbit torque magnetic random access memory (SOT-MRAM) [5-8] due to its low power consumption and efficient magnetization switching. A charge current density $j_C$ in a heavy metal (HM) is converted into a spin current density $j_S$ via SHE, which then exerts an in-plane damping-like SOT $\tau_{DL}$ on the magnetization of the adjacent ferromagnet (FM) [9, 10]. The ratio of $j_S$ to $j_C$ is called the damping-like torque (DLT) efficiency $\theta_{DL}$ (also termed as charge-to-spin conversion efficiency or spin Hall angle). With the strength of spin-orbit coupling (SOC) depending on the atomic number Z, 5d transition metals such as Pt, Ta, W [9, 11-15] have been improved by alloying with other heavy metals with large Z such as Au, Pd [16-18]. Alternatively, non-metallic elements (impurities with smaller Z) have been incorporated into 5d transition metals (host with larger Z). The difference in Z between the host and the impurity has been found to result in an enhancement of SHE [19]. $\theta_{DL}$ has been enhanced by incorporating non-metallic elements into 5d transition metals such as sulfur (S) in Pt [20], oxygen (O) in Pt [21-24], Ta [25], and W [26], and nitrogen (N) in Pt [27], Ta [28] and W [29]. However, the effects of nitrogen (N) incorporation are still underexplored, especially in controlling the longitudinal resistivity $\rho_{xx}$, which is an important yardstick to be considered for SOT-MRAM applications. Incorporation of nitrogen has led to an undesirable increase in $\rho_{xx}$ of Ta(N) and W(N) [28, 29]. Furthermore, the origin responsible for the enhancement in SHE in non-metallic doped Pt, Ta, and W has not yet been confirmed explicitly [25, 27, 28, 29]. Studies on the temperature dependence of the SHE in 5d transitions metal having non-metallic impurities, are scarce. Most room temperature studies have reported a limited variation of $\theta_{DL}$ with resistivity $\rho_{xx}$.

There have been extensive efforts to enhance the SHE, mainly via two mechanisms:

intrinsic and extrinsic SHE [30-34]. The intrinsic SHE depends on the berry curvature of the material, in which an anomalous velocity arises from a momentum-space berry phase [31]. It leads to an elastic event in which the wave-vector $\vec{k}$ of the up-spin and down-spin electrons generated from charge current is conserved [32, 33], and is typically seen in 4d and 5d transition metals. The extrinsic SHE arises when impurities are introduced in the HM, and can be further classified into side-jump and skew-scattering [34]. For side-jump scattering, a discontinuous side-ways displacement is created near the impurities for the up-spin and down-spin electrons generated from charge current, leading to an elastic event due to the cancellation of $\vec{k}$, and is found in materials with high amount of impurity concentrations in host. The skew scattering, however, is different from the intrinsic and side-jump mechanism as scattering bends or skews the trajectories of up-spin and down-spin electrons in different directions, and is found in super-clean or low resistivity materials [32]. This leads to a condition where $\vec{k}$ is not conserved, resulting in an inelastic event. Given the separation of SHE based on elastic and inelastic events, a strong correlation between spin Hall conductivity, $\sigma_{SH}^{xy}$ and momentum relaxation time $\tau_{relax}$ can be obtained. Intrinsic SHE and side-jump scattering share the same scaling, with $\sigma_{SH}^{xy}$ being independent of $\tau_{relax}$. Skew scattering shows $\sigma_{SH}^{xy} \propto \tau_{relax}$ scaling. Therefore, it is hard to disentangle the contribution of the intrinsic from the side-jump scattering. Despite tremendous efforts [18, 27, 28, 29], a clear separation between the contributions of intrinsic and side-jump scattering to the SHE in non-metallic element doped HM has eluded us so far. The limited variation of $\theta_{DL}$ with resistivity $\rho_{xx}$ at room temperature and the choice of host/impurity combinations, especially for incorporation of non-metallic elements in the HM, need to be addressed.

In this paper, we present a successful disentanglement of intrinsic and extrinsic side-jump scattering by studying the SHE using a non-metallic Nitrogen (N) implanted in Pt, at 100-293 K using spin-torque ferromagnetic resonance (ST-FMR) lineshape (spectral) analysis. We observe a crossover from intrinsic to extrinsic side-jump scattering mechanism as the N-ion dose increases from $2\times10^{16}$ to $1\times10^{17}$ ions/cm$^2$. This happens due to a decrease in intrinsic spin Hall conductivity, $\sigma_{SH}^{int}$ which is counterbalanced by an increase in side-jump induced SHE efficiency, $\theta_{SH}^{sj}$. Our results offer an interesting opportunity to understand the underlying SHE mechanism for incorporation of nonmetallic elements in HM, by use of a novel approach of ion-implantation.

## II. EXPERIMENTAL DETAILS

Thin films of Pt (10 nm)/ MgO (10 nm)/ $Al_2O_3$ (10 nm) layers were deposited on a Si/$SiO_2$ substrate at room temperature using an ultrahigh vacuum sputtering. The thin film stacks were implanted sequentially with doses of $2\times10^{16}$ ions/cm$^2$, $5\times10^{16}$ ions/cm$^2$ and $1\times10^{17}$ ions/cm$^2$ by an N-ion source beam having an energy of 20 keV. After ion implantation, the capping layers of MgO and $Al_2O_3$ were removed by $Ar^+$ ion milling and then a FM layer of NiFe (5nm) was sputtered on these samples (See Appendix A). Hereafter, they will be referred to as Pt(N) $2\times10^{16}$, Pt(N) $5\times10^{16}$, and Pt(N) $1\times10^{17}$. All the three bilayer samples were then patterned into rectangular micro strips using photolithography. Thereafter, Ti (10 nm)/Al (200 nm) electrodes were deposited. The design of the co-planar waveguides for ST-FMR measurements is shown in Fig. 1(a). The temperature dependent ST-FMR measurements were performed in the range of 100-293 K.

## III. RESULTS AND DISCUSSION

### A. Dose-dependent ST-FMR measurements at room temperature (293 K)

To study the influence of dose on SHE in Pt(N), we first performed the ST-FMR based lineshape analysis to determine $\theta_{DL}$. Figure 1(a) shows the schematic of the ST-FMR measurement set-up. In this technique, when a microwave current $I_{rf}$ flows in the longitudinal direction of HM/FM bilayer, a transverse spin current density $j_s$ is generated, which exerts a DLT, $\tau_{DL}$ on the local magnetization of FM. So, an $I_{rf}$ was passed in the longitudinal direction of the device with an applied power of 10 dBm. The $I_{rf}$ generates an rf Oersted field $h_{rf}$ (according to Ampère's Law), which simultaneously exerts an Oersted field torque $\tau_{OFT}$. An external magnetic field $\mu_o H_{ext}$ was swept in the range of ±240 mT at an angle of $\phi = 45^o$ with respect to the longitudinal direction of the device. At resonance condition, both $\tau_{DL}$ and $\tau_{OFT}$ drive the magnetization precession in the FM, which results in a periodically varying resistance ΔR due to the anisotropic magnetoresistance (AMR) of NiFe. The mixing of the oscillating ΔR and $I_{rf}$ produces a DC voltage $V_{mix}$, which is detected using a lock-in amplifier via a bias tee, and is expressed as [9, 20, 25, 26]:

$$V_{mix} = SF_{sym}(H_{ext}) + AF_{asym}(H_{ext}), \tag{1}$$

where, $F_{sym}(H_{ext}) = \frac{(\Delta H)^2}{(H_{ext}-H_o)^2+(\Delta H)^2}$, is the symmetric part of the $V_{mix}$ spectrum, $F_{asym}(H_{ext}) = \frac{\Delta H(H_{ext}-H_o)}{(H_{ext}-H_o)^2+(\Delta H)^2}$, is the antisymmetric part, ΔH and $H_0$ are the half-width-at-half-maximum (linewidth) and the resonance field, and S and A are the weight factors of the symmetric and antisymmetric spectra respectively. For the observed spectra, while the symmetric component is dominated by the $\tau_{DL}$ contribution (from $j_S$), the antisymmetric component is primarily dominated by $\tau_{OFT}$ (from $j_C$). Figure 1(b) shows the de-convoluted ST-FMR spectra of $V_{mix}$ measured at 5 GHz for Pt(N) $5\times10^{16}$. Please refer to Appendix B to see the de-convoluted ST-FMR spectra obtained for other samples. The broad range of ST-FMR spectra obtained for applied frequency f = 5-11 GHz is shown for all the samples (Appendix B). The Gilbert damping parameter α which depends on linewidth ΔH, is estimated using [21]:

$$\Delta H = \Delta H_o + \frac{2\pi f}{\gamma}\alpha, \tag{2}$$

where, γ is the gyromagnetic ratio and $\Delta H_o$ is the inhomogeneous linewidth broadening which is independent of f. Referring to Appendix B (Fig. 6(f)), $\alpha$ is estimated from the slope of $\Delta H$ plotted as a function of f. The value of $\alpha$ is higher for Pt(N) $5\times10^{16}$ as compared to that of Pt(N) $2\times10^{16}$ and Pt(N) $1\times10^{17}$ . To quantify $\theta_{DL}$, we performed the lineshape analysis of the ST-FMR spectrum, using Eq. (3) [9]:

$$\theta_{DL} = \frac{S}{A}\frac{e\mu_o M_s td}{\hbar}\sqrt{1+\frac{M_{eff}}{H_o}}, \tag{3}$$

where e is the elementary charge, ħ is the reduced Planck constant, t is the thickness of the NiFe layer, d is the thickness of the heavy metal layer and the effective magnetization $M_{eff}$ is obtained from Kittel fitting. The figures in Appendix B show the obtained values of $\theta_{DL}$ for the studied frequency range of 5-11 GHz. $\theta_{DL}$ is found to be invariant with frequency, implying a negligible role of thermal effect and non-controlled relative phase between $I_{rf}$ and $h_{rf}$ that arises from sample design [21, 35, 36]. The average $\theta_{DL}$ values obtained are 0.119 ± 0.002 for Pt(N) $2\times10^{16}$, 0.132 ± 0.008 for Pt(N) $5\times10^{16}$ and 0.098 ± 0.008 for Pt(N) $1\times10^{17}$. Noticeably, the $\theta_{DL}$ of pure sample (Pt/NiFe) is found to be 0.062± 0.004. It demonstrates that ion implantation provides a better alternative to incorporate nitrogen in Pt when compared to the sputtering method, as even a small dose of $2\times10^{16}$ ions/cm$^2$ in Pt leads to ~1.9 times enhancement in $\theta_{DL}$ from 0.062 to 0.119. We find a ~2.2 times enhancement in $\theta_{DL}$ from 0.062 (Pt) to 0.132 (Pt(N) $5\times10^{16}$ ). However, we observe a non-monotonic dependence of $\theta_{DL}$ on implantation dose, similar to Xu et. al. [27] where nitrogen was incorporated in Pt via sputtering. On the contrary, we recently observed a monotonic dependence of $\theta_{DL}$ on oxygen (O) implantation dose [21]. We also performed the angular dependent ST-FMR measurements (See Appendix C). The unbroken two-fold and mirror symmetries of the torques along with the negligible spin-pumping (See Appendix D) contribution allows us to use the lineshape analysis to quantify the $\theta_{DL}$.

To picturize the correlation among $\rho_{xx}$, $\alpha$, $\mu_o M_{eff}$, spin mixing conductance $g_{eff}^{\uparrow\downarrow}$ and $\theta_{DL}$ for Pt(N), we plot these parameters as a function of implantation dose as shown in Fig. 2. First, $\rho_{xx}$ is found to be monotonically increasing in Fig. 2(a), which is similar to previous report [27]. Second, $\alpha$ shows a non-monotonic dependence with an increase of N-ion dose, where a maximum value of 0.032 is obtained for the dose of $5\times10^{16}$ (Fig. 2(b)). Third, $M_{eff}$ decreases with increasing N-ion dose (Fig. 2(c)). Noticeably, a minimum $\mu_o M_{eff}$ = 610 mT is seen for Pt(N) $5\times10^{16}$ as compared to 711 mT for Pt(N) $2\times10^{16}$ and 686 mT for Pt(N) $1\times10^{17}$, indicating a change in the perpendicular magnetic anisotropy field $H_p$. The $\mu_o M_{eff}$ of pure Pt is found to be 765 mT, indicating less $H_p$. Fourth, the $g_{eff}^{\uparrow\downarrow}$ shows a similar trend as obtained for $\alpha$ (Fig. 2(d)), indicating an enhanced $j_S$ at the HM/FM interface. Fifth, $\theta_{DL}$ in Fig. 2(e) shows a similar trend as obtained for $\alpha$ and $g_{eff}^{\uparrow\downarrow}$. Summarizing the dose dependent results, both $\alpha$ and $\theta_{DL}$ increase monotonically from 0 to $5\times10^{16}$ ions/cm$^2$ and then suddenly decrease for $1\times10^{17}$ ions/cm$^2$, similar to the results obtained by Xu et. al. [27]. In agreement with the highest $\alpha$ and $g_{eff}^{\uparrow\downarrow}$ for Pt(N) $5\times10^{16}$, $\theta_{DL}$ is found to be maximum for Pt(N) $5\times10^{16}$ [37]. However, due to the limited variation of $\theta_{DL}$ with $\rho_{xx}$, studying the $\theta_{DL}$ for different doses of impurities at room temperature alone may be insufficient in understanding the underlying mechanism. Hence, it is also important to investigate the dependence of $\theta_{DL}$ and associated properties on temperature.

**B. Temperature-dependent ST-FMR measurements.**

In order to gain a deeper understanding of the enhancement in SHE of Pt(N), we performed temperature (T) dependent ST-FMR measurements in the range of 100-293 K. A linear increase of $\rho_{xx}$ as a function of T (for T = 10-293 K) is observed in Fig. 3(a), which confirms a metallic behavior [31]. Fitting a straight line to the data and extrapolating to T = 0 K allowed us to deduce the residual resistivity $\rho_{xx,0}$, which is summarized in Table II. Second, the linewidth $\Delta H$ and gilbert damping parameter $\alpha$ increase at lower T as shown in Fig. 3 (b. i, ii). This enhancement in $\alpha$ may be due to two reasons: an increase of $j_S$ at lower T, and the increase of magnetic damping at the surface contribution. We consider the latter to be unlikely as it arises in a ferromagnet at a very low temperature range and with low thickness [38]. Third, $\mu_o M_{eff}$ increases at lower T, as seen in Fig. 3(c). This is in accordance with $M_s \propto \frac{1}{T}$, assuming that $H_p$ remains invariant with temperature [39]. Fourth, to confirm the high $j_S$ created at FM/HM interface [37], $g_{eff}^{\uparrow\downarrow}$ (See Appendix E) is found to increase with decreasing T, as shown in Fig. 3(d), especially for the higher doses of Pt(N) $5\times10^{16}$ and Pt(N) $1\times10^{17}$. Lastly, $\theta_{DL}$ is plotted as a function of T in Fig. 3 (e), and is found to increase with decreasing T for Pt(N) $5\times10^{16}$ and Pt(N) $1\times10^{17}$. Consequently, a high $\theta_{DL}$ of 0.18-0.19 is obtained for Pt(N) $5\times10^{16}$ and Pt(N) $1\times10^{17}$. The $\theta_{DL}$ is found to be invariant with T for Pt(N) $2\times10^{16}$ and Pt. Such a kind of increase/decrease of $\theta_{DL}$ with T hints at the possibility of an intrinsic and/or extrinsic side-jump contribution [33, 40, 41].

### C. Contribution to SHE

In anomalous Hall effect (AHE), the side-jump term, $\rho_{sj}$ or $\sigma_{sj}$ (proposed by Berger) [42] arising from extrinsic effect was confusingly viewed as an intrinsic term, $\rho_{int}$ or $\sigma_{int}$ (proposed by Karplus and Luttinger) [43]. This happened due to both the $\rho_{int}$ and $\rho_{sj}$ being proportional to $\rho_{xx}^2$ (or, simply $\sigma_{int} \propto \sigma_{xx}^2$ and $\sigma_{sj} \propto \sigma_{xx}^2$), where $\rho_{xx}^2$ is square of resistivity and $\sigma_{xx}^2$ is square of conductivity. This was accepted until the concept of residual resistivity arising from impurities, $\rho_{xx,0}$ was introduced by Tian et al., [44]. They proposed that the proper scaling for the AHE should involve not only the $\rho_{xx}$ (or $\sigma_{xx}$), but also an important term, the residual resistivity $\rho_{xx,0}$. The AHE and SHE share the same analogy as demonstrated experimentally by Moriya et. al. [45].

Therefore, the total spin Hall conductivity $\sigma_{SH}^{xy}$ can be expressed as a sum of intrinsic and extrinsic SHE (side-jump and skew scattering) [30, 33, 44, 46]:

$$\left|\sigma_{SH}^{xy}\right| = \sigma_{SH}^{int} + \left|\sigma_{SH}^{sj} + \sigma_{SH}^{ss}\right|, \quad (4)$$

$$= \sigma_{SH}^{int} + \left|(\theta_{SH}^{sj}\,\rho_{xx,0} + \theta_{SH}^{ss}\,\rho_{xx,0})\,\sigma_{xx}^{2}\right|, \quad (5)$$

where, $\sigma_{SH}^{int}$ is intrinsic spin Hall conductivity, $\sigma_{SH}^{sj}$ is spin Hall conductivity due to side-jump scattering, $\sigma_{SH}^{ss}$ is spin Hall conductivity due to skew scattering, $\theta_{SH}^{sj}$ is side-jump induced SHE efficiency, $\theta_{SH}^{ss}$ is skew scattering induced SHE efficiency, and $\sigma_{xx}$ is conductivity.

To elucidate the explicit contribution from intrinsic and side-jump to the SHE, and to understand if skew scattering has any significant role to play, $\sigma_{SH}^{xy}$ is plotted as a function of $\sigma_{xx}$ in the inset of Fig. 4(a). $\sigma_{xx}$ is in the range of $10^4\ \Omega^{-1}cm^{-1}$. In the analogy to AHE, skew scattering arises in the higher conductivity range of super clean metals ($10^6 < \sigma_{xx} < 10^8\ \Omega^{-1}cm^{-1}$) [47]. Additionally, impurity induced skew scattering shows a T-independent $\theta_{DL}$ which was not observed for the higher doses in our samples [34, 40, 41]. Hence, skew scattering is not a possible mechanism in our samples. After excluding skew scattering, Eq. (5) can be expressed as:

$$\left|\sigma_{SH}^{xy}\right| = \sigma_{SH}^{int} + \left|(\theta_{SH}^{sj}\,\rho_{xx,0})\sigma_{xx}^{2}\right|, \quad (6)$$

To probe the exact contributions from intrinsic and side-jump scattering, $\sigma_{SH}^{xy}$ is plotted as a function of $\sigma_{xx}^{2}$ in Fig. 4(a). Using the value of $\rho_{xx,0}$, $\sigma_{SH}^{int}$ of 1303.29 $(\frac{\hbar}{2e})\Omega^{-1}cm^{-1}$ is obtained for Pt, which is found to be very close to the theoretical value of 1300 $(\frac{\hbar}{2e})\ \Omega^{-1}cm^{-1}$ reported by Tanaka et. al., [48]. Further, for Pt(N), $\sigma_{SH}^{xy}$ is fitted to Eq. (6) and a $\sigma_{SH}^{int}$ of 1283.55 $(\frac{\hbar}{2e})\Omega^{-1}cm^{-1}$ is obtained for lower dose Pt(N) $2\times10^{16}$ as shown in Fig. 4(b) . This is close to $\sigma_{SH}^{int}$ of Pt, hinting at the dominant intrinsic SHE mechanism. However, the surprising result is the lower $\sigma_{SH}^{int}$ of 1146.01 $(\frac{\hbar}{2e})\Omega^{-1}cm^{-1}$ and 422.72 $(\frac{\hbar}{2e})\Omega^{-1}cm^{-1}$ for Pt(N) $5\times10^{16}$ and Pt(N) $1\times10^{17}$, respectively. A sudden decrease in $\sigma_{SH}^{int}$ is counterbalanced by an increase of $\theta_{SH}^{sj}$. The $\theta_{SH}^{sj}$ is found to be 0.31 and 0.26 for Pt(N) $5\times10^{16}$ and Pt(N) $1\times10^{17}$, respectively (Fig. 4 (c)). Pt is a spin Hall material having a positive SHE sign [9, 11, 48, 49] and so, the positive sign of $\sigma_{SH}^{int}$ indicates that intrinsic SHE still has some contribution to the SHE [30, 33, 47, 49]. Please see table II for details. Hence, the increase in $\theta_{DL}$ in Pt (N) is also influenced by extrinsic side-jump scattering, especially for the high implanted dose samples Pt(N) $5\times10^{16}$ and Pt(N) $1\times10^{17}$. Therefore, with the increase in dose, we observe an increase in extrinsic side-jump contribution to SHE, which could play a significant role in the enhancement of SHE. This also leads to a reduction in the intrinsic SHE. Therefore, a crossover of intrinsic to extrinsic side-jump induced SHE is observed as the implantation of N-dose from $2\times10^{16}$ ions/cm$^2$ to $1\times10^{17}$ ions/cm$^2$ is increased in the Pt layer. The successful disentanglement of intrinsic and extrinsic side jump induced by SHE could be a promising approach to understand the mechanism for enhancement in SHE.

## 4. CONCLUSION

In conclusion, we utilized a novel approach of ion-implantation to incorporate Nitrogen in Pt by varying dose from $2\times10^{16}$ to $1\times10^{17}$ ions/cm$^2$. We studied the dependence of $\theta_{DL}$ on both N-ion dose and temperature, to disentangle the intrinsic and extrinsic side-jump scattering mechanism in spin Hall effect. We found a crossover of intrinsic to extrinsic side-jump induced SHE as N-ion dose increased from $2\times10^{16}$ ions/cm$^2$ to $1\times10^{17}$ ions/cm$^2$. A sudden decrease in $\sigma_{SH}^{int}$ is counterbalanced by the increase in $\theta_{SH}^{sj}$. These results indicate that studying the $\theta_{DL}$ as a function of implantation dose, as well as a function of temperature, is important to understand the underlying physical phenomenon contributing to SHE. We believe that such a deep comprehension of enhancement in SHE may help us in revealing the host and impurity combination to unlock the full potential of SHE in 5d transition metals for SOT-MRAM application.

**Data availability statement**

The data that support the findings of this study are available from the corresponding author upon reasonable request.

**Acknowledgements**

U.S. would like to acknowledge the Kanazawa Memorial Foundation Scholarship. Y.F. and H.A. would like to acknowledge JSPS Grant-in-Aid (KAKENHI No. 18H01862 and 19K21112) and KIOXIA Corporation. R.M., S.G. and R.S.R. would like to acknowledge research grants NRF-CRP21-2018-0003 and MOE2019-T2-1-058 from National Research Foundation and Ministry of Education Singapore, respectively. The authors thank Mr. Raj Kumar, IUAC for his support in Nitrogen ion implantation using the Tabletop Accelerator.

Table I: $\boldsymbol{\alpha}$, $\mathbf{M_{eff}}$ , $\mathbf{g_{eff}^{\uparrow\downarrow}}$, $\boldsymbol{\theta_{DL}}$, and $\boldsymbol{\rho_{xx}}$ for different N-ion dose

| **Sample** | $\boldsymbol{\alpha}$ | $\boldsymbol{\mu_o}\mathbf{M_{eff}}$ (mT) | $\mathbf{g_{eff}^{\uparrow\downarrow}}$ $(\mathbf{10^{19} m^{-2}})$ | $\boldsymbol{\theta_{DL}}$ | $\boldsymbol{\rho_{xx}}$ $(\boldsymbol{\mu\Omega}$ **cm**$)$ |
|---|---|---|---|---|---|
| Pure Pt | 0.012 | 765 | 1.91 | 0.062 | 18.8 |
| Pt(N) $2\times10^{16}$ ions/cm$^2$ | 0.022 | 711 | 2.08 | 0.119 | 39.7 |
| Pt(N) $5\times10^{16}$ ions/cm$^2$ | 0.032 | 610 | 3.10 | 0.132 | 65.3 |
| Pt(N) $1\times10^{17}$ ions/cm$^2$ | 0.024 | 686 | 2.57 | 0.098 | 146.3 |

Table II: $\boldsymbol{\rho_{xx,0}}$, $\boldsymbol{\sigma_{SH}^{int}}$ , and $\boldsymbol{\theta_{SH}^{sj}}$ for different N-ion dose

| **Sample** | $\boldsymbol{\rho_{xx,0}}$ $(\boldsymbol{\mu\Omega}$ **cm**$)$ | $\boldsymbol{\sigma_{SH}^{int}}$ $(\frac{\hbar}{2e}\boldsymbol{\Omega^{-1}}\mathbf{cm^{-1}})$ | $\boldsymbol{\theta_{SH}^{sj}}$ |
|---|---|---|---|
| Pure Pt | 9.01 | 1303.29 | 0.069 |
| Pt(N) $2\times10^{16}$ ions/cm$^2$ | 14.8 | 1283.55 | 0.251 |
| Pt(N) $5\times10^{16}$ ions/cm$^2$ | 31.1 | 1146.01 | 0.315 |
| Pt(N) $1\times10^{17}$ ions/cm$^2$ | 78.8 | 422.72 | 0.262 |

**Figure Captions:**

**Figure** 1. **ST-FMR measurement setup:** (a) Schematic showing ST-FMR measurement technique and detection principle for a bilayer thin film, along with an optical image of the micro-device. (b) ST-FMR spectra ($V_{mix}$) for f = 5 GHz obtained for Pt(N) $5\times10^{16}$, fitted using Eq. (1). De-convolution fitting of $V_{mix}$ into symmetric and antisymmetric components displayed by brown and violet solid lines respectively.

**Figure** 2. **Dose-dependent ST-FMR measurements:** (a) $\rho_{xx}$, (b) $\alpha$, (c) $\mu_o M_{eff}$, (d) $g_{eff}^{\uparrow\downarrow}$, and (e) $\theta_{DL}$ for different doses of nitrogen in Pt, at room temperature (293 K).

**Figure** 3. **T-dependent ST-FMR measurements:** (a) $\rho_{xx}$, (b. i) $\Delta H$, (b. ii) $\alpha$, (c) $\mu_o M_{eff}$, (d) $g_{eff}^{\uparrow\downarrow}$, and (e) $\theta_{DL}$ for different doses of nitrogen in Pt plotted as a function of temperature. The solid line in Fig. 3 (a) represents the linear fitting.

**Figure** 4. **Contribution to SHE**: (a) $\sigma_{SH}^{xy}$ plotted as a function of $\sigma_{xx}^2$ for implantation (orange data points) and Pure Pt (black data points). The solid lines represent the fitting using Eq. (6). Inset shows $\sigma_{SH}^{xy}$ plotted as a function of $\sigma_{xx}$. (b) $\sigma_{SH}^{int}$, and (c) $\theta_{SH}^{sj}$ plotted as a function of N-ion dose.

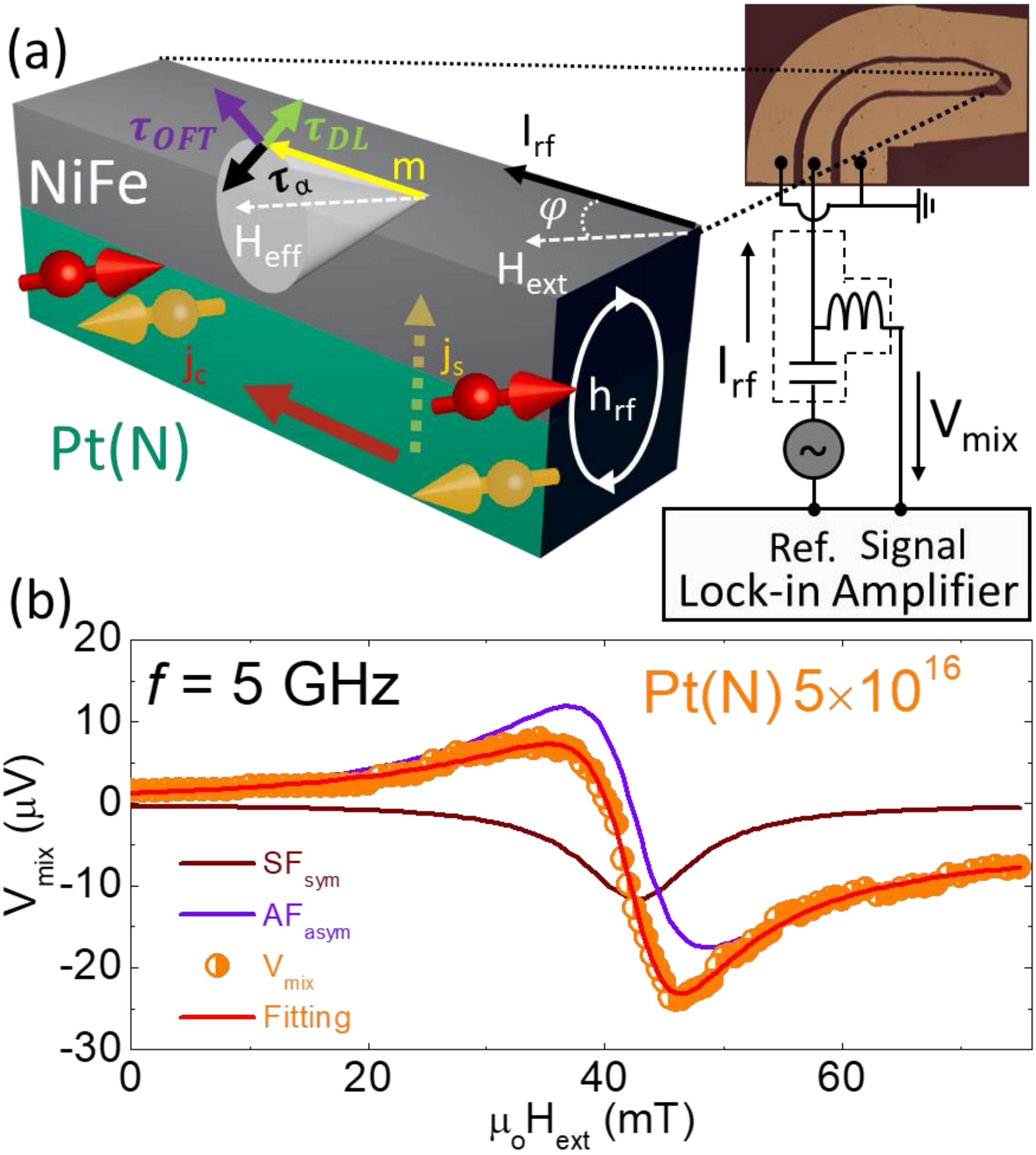

Fig. 1. U. Shashank *et. al.*

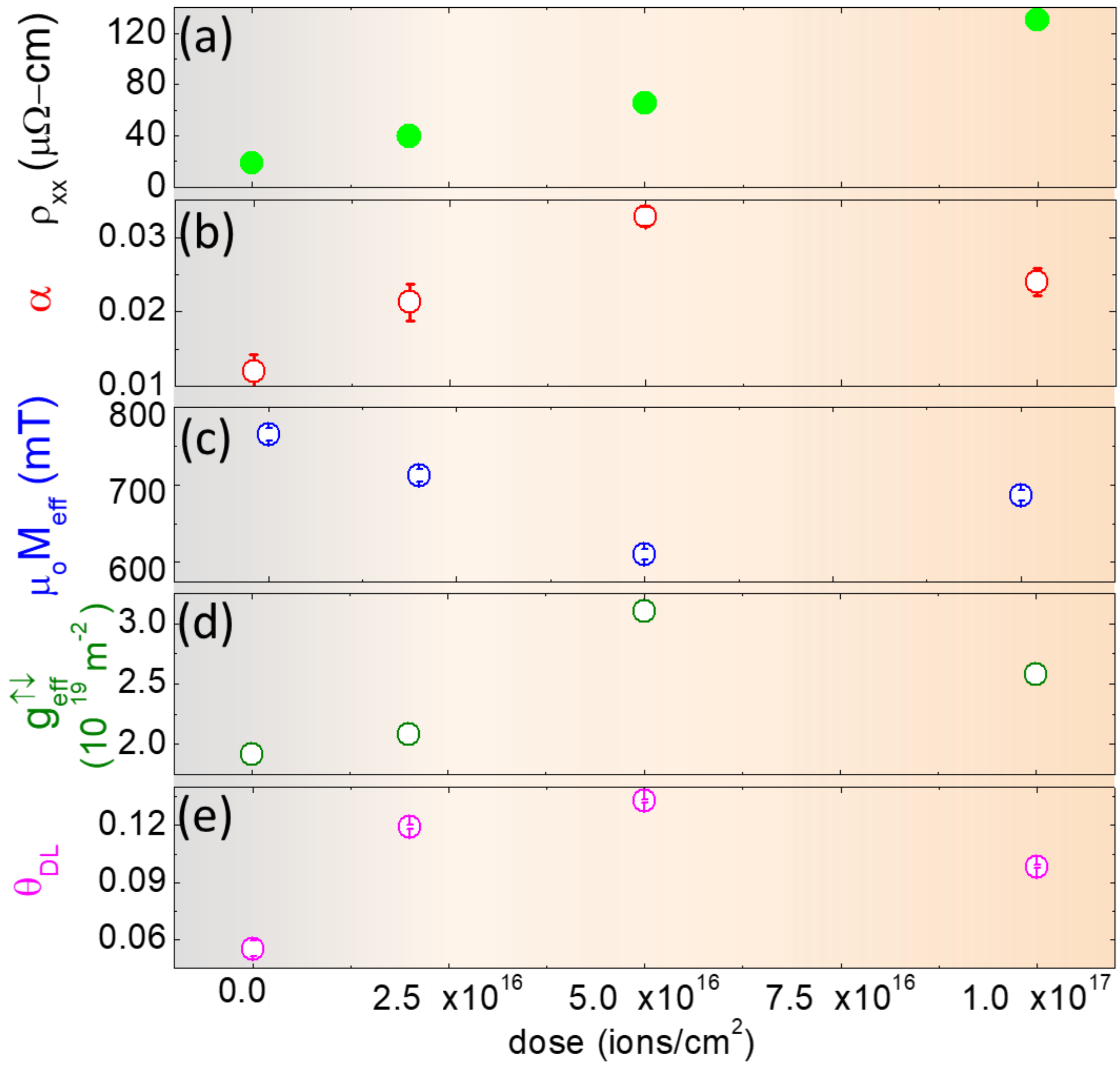

Fig. 2. U. Shashank *et. al.*

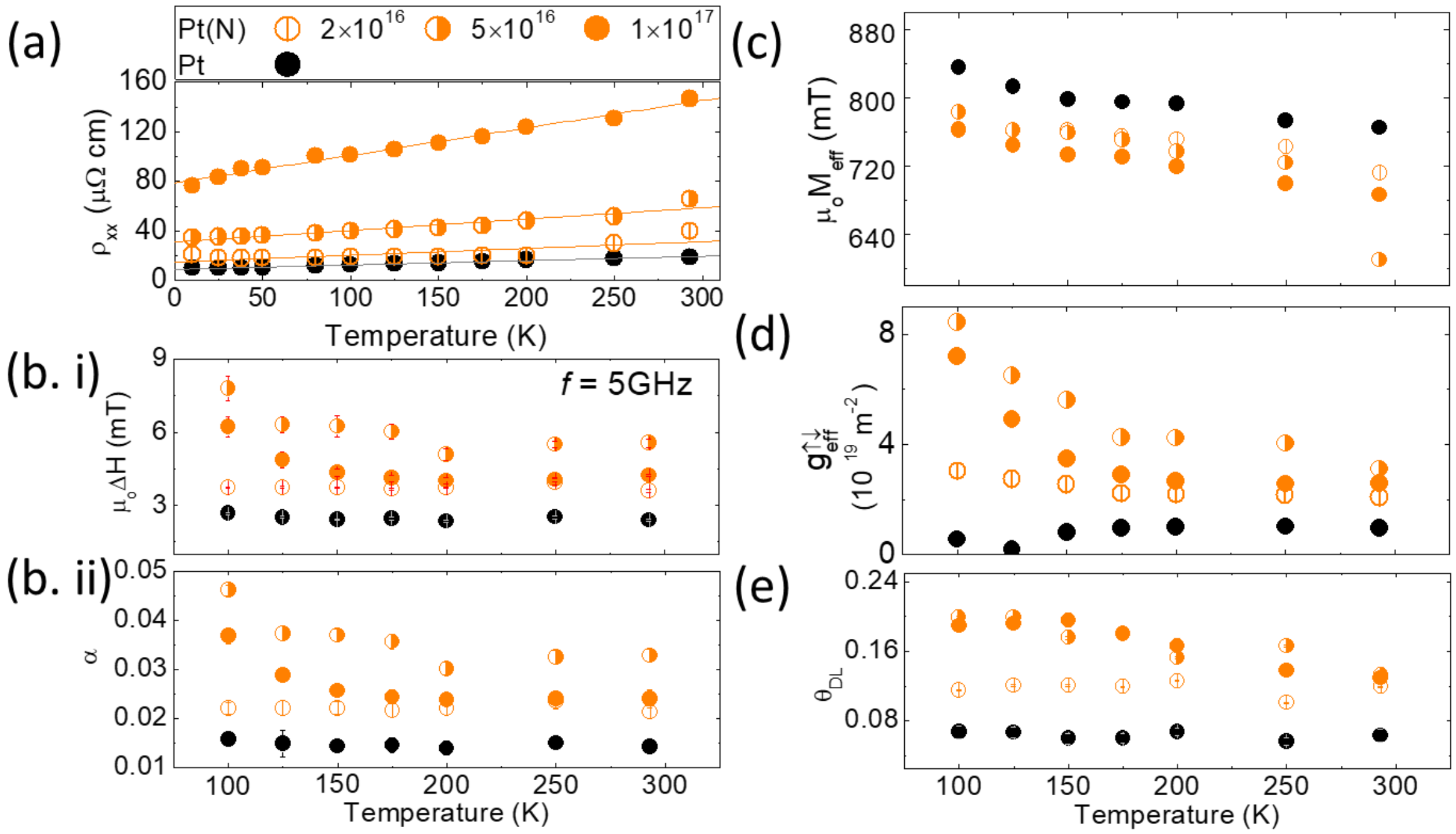

Fig. 3. U. Shashank *et. al.*

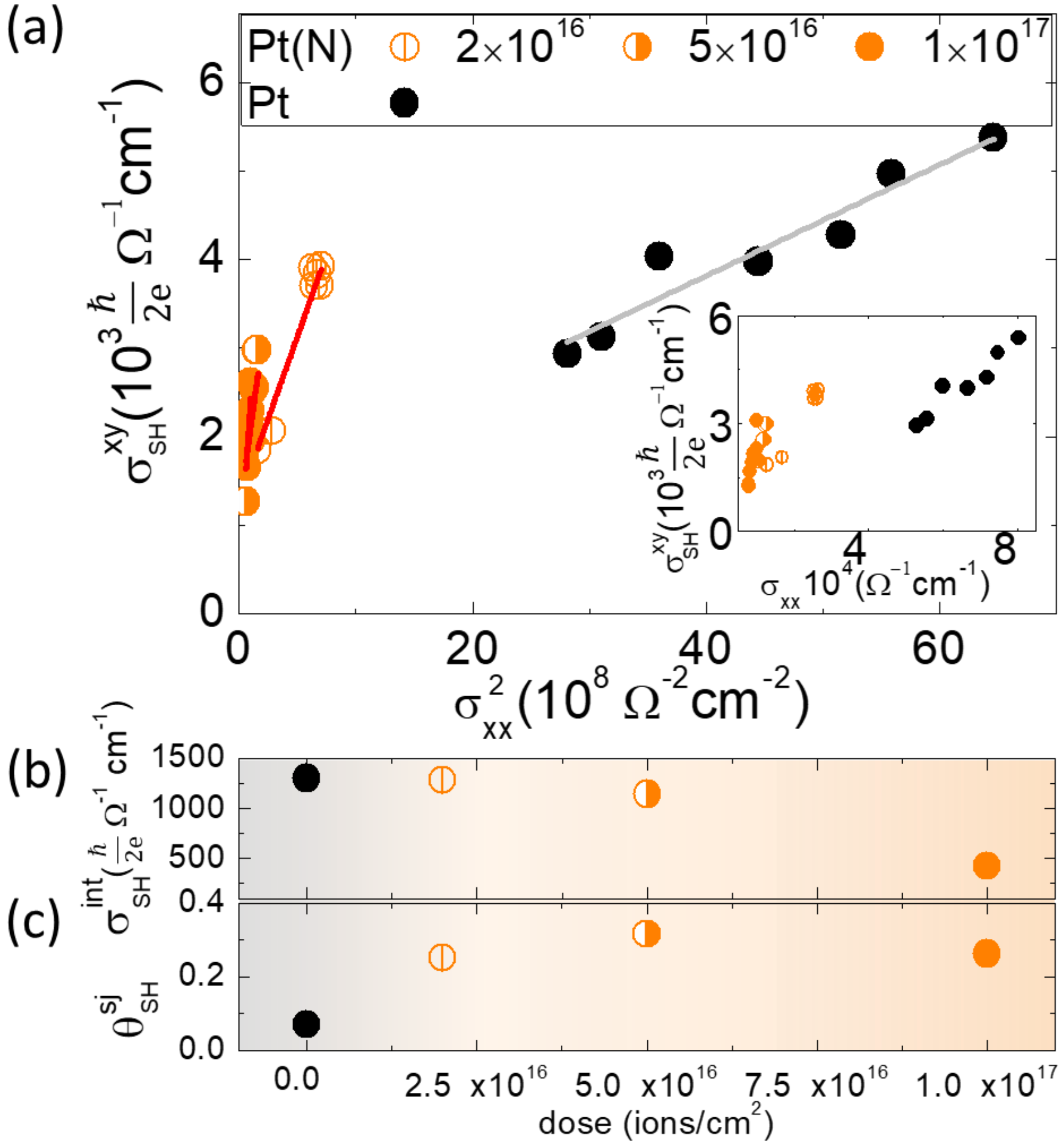

Fig. 4. U. Shashank *et. al.*

## APPENDIX A: ION IMPLANTATION METHOD

The present experiment was carried out using a 30 kV ion accelerator/ implanter installed at the Inter-University Accelerator Centre (IUAC), New Delhi. This system comprises of an ion source, a cold plasma-based Penning ion generator, assembled in a nylon housing and connected to a 30 kV power supply which is capable of delivering a stable current of up to 350 μA. An Einzel lens and an electrostatic quadrupole are used for focusing the ion beams. The required ions are selected by adjusting the magnetic field strength of a bending magnet which provides a uniform, variable field of 0.21-0.35 T. The ion beam spot size of $15 \times 15\ \mathrm{mm}^2$ can be scanned upon the target in the implantation chamber having a vacuum of $1.3 \times 10^{-4}$ Pa.

The ion dose of fluence is defined as the number of ions implanted per unit area of the target. The number of ions is determined by allowing the beam to hit the inner wall of a Faraday cup (Conductive) and measuring the resulting current by a current Integrator. Hence, the dose or fluence (ions /cm$^2$) is calculated by using the relation:

$$\mathrm{Dose} = \frac{\int \mathrm{I\,dt}}{\mathrm{qe(Area)}}\ (\mathrm{ions/cm^2}), \tag{7}$$

where I is the beam current, t is the time, q is the charge state of the ion, e is the electronic or elementary charge, and Area is the area of the target.

The deposited multilayer stack of Pt (10 nm)/ MgO (10 nm)/ $Al_2O_3$ (10 nm) were implanted with 20 keV N ion beam, as shown in Fig. 5 (a). The two protective layers (MgO and $Al_2O_3$) were deposited above Pt for the following two reasons: First, the protective layers allowed the least disturbance to the target Pt layer while ensuring the uniform distribution of N ions in Pt following a similar recipe of our previous work [20]. Second, the MgO provides a good end-point detector signal during the $Ar^+$ ion milling whereas $Al_2O_3$ protects the hygroscopic MgO from self-sputtering in the implantation process [20]. Then, oxide capping layers of MgO and $Al_2O_3$ were removed by ion milling (Fig. 5 (b)) and top FM layer of NiFe (5nm) was then sputtered on these samples (Fig. 5 (c)).

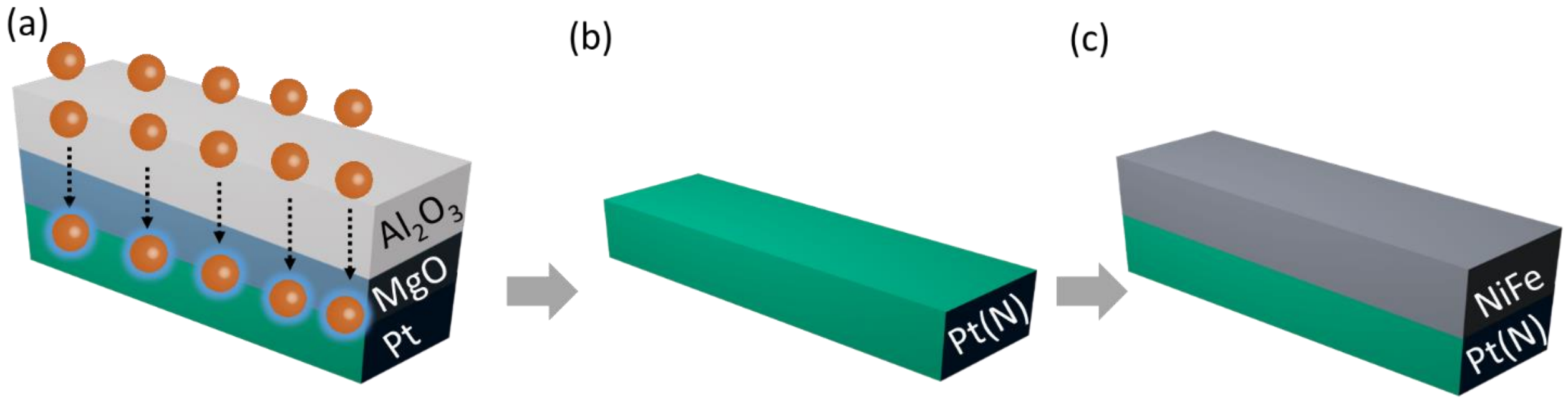

Fig. 5. **Ion implantation method:** (a) Visualization of ion implantation followed by removal of protective layers (b) and (c) after the deposition of NiFe on top of Pt.

## APPENDIX B: LINESHAPE ANALYSIS AT 293 K (ROOM TEMPERATURE)

We performed the ST-FMR based lineshape analysis in a wide range of frequency f= 5-11 GHz to extract the symmetric and antisymmetric components from $V_{mix}$, i.e. $F_{sym}(H_{ext})$ and $F_{asym}(H_{ext})$, multiplied with the weight factors S and A. From $V_{mix}$, linewidth ($\mu_o\Delta H$) and S/A are used to calculate the α and $\theta_{DL}$ respectively.

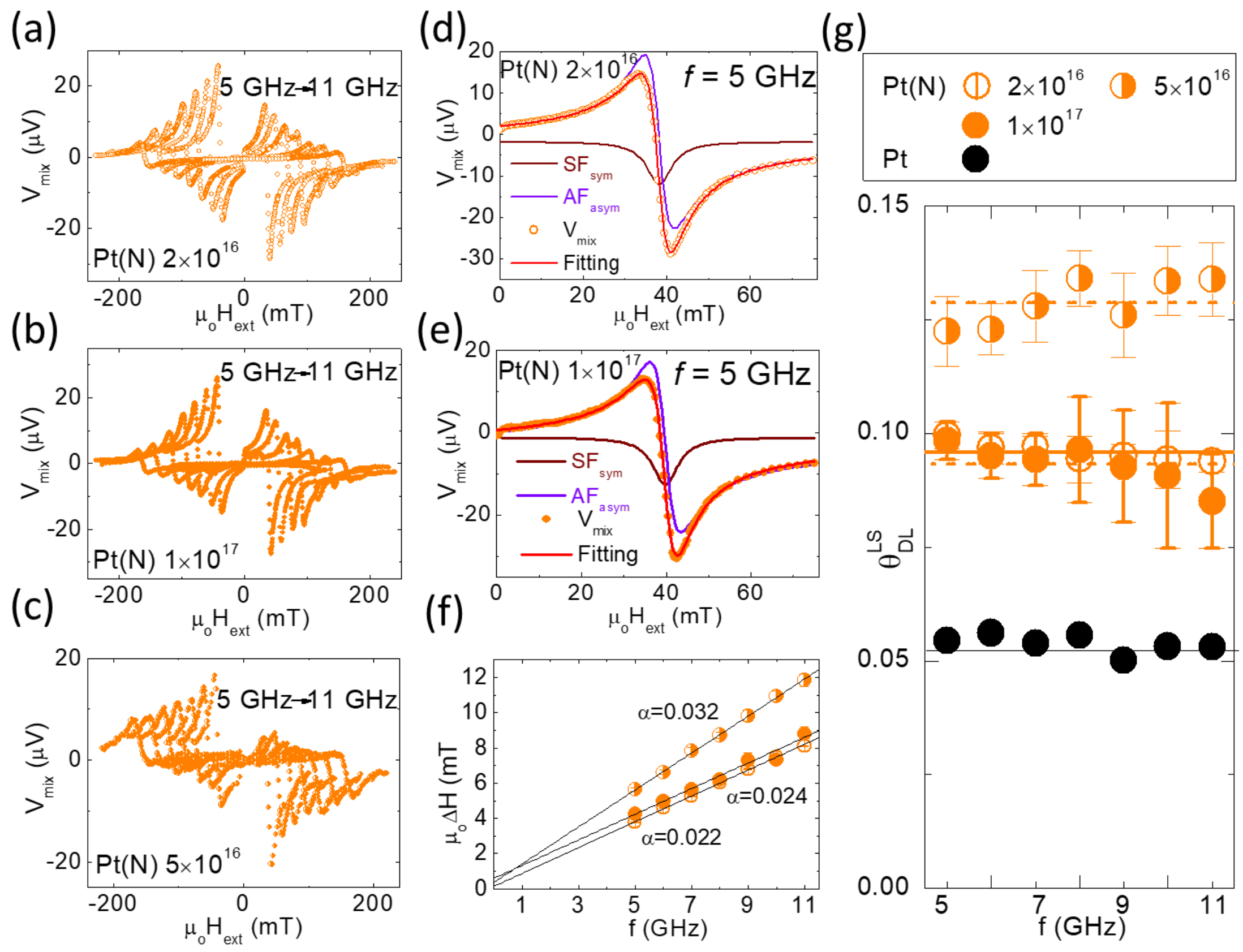

Fig. 6. ST-FMR spectra $V_{mix}$ obtained for (a) Pt(N) $2\times10^{16}$, (b) Pt(N) $1\times10^{17}$, (c) Pt(N) $5\times10^{16}$. De-convolution fitting of $V_{mix}$ measured at f = 5 GHz into symmetric and antisymmetric components displayed by brown and violet solid lines respectively for (d) Pt(N) $2\times10^{16}$, and (e) Pt(N) $1\times10^{17}$, (f) $\mu_o\Delta H$ vs f (with solid lines as linear fit), and (g) Frequency invariant $\theta_{DL}$ obtained for Pt(N) $2\times10^{16}$, Pt(N) $5\times10^{16}$, Pt(N) $1\times10^{17}$ and Pure Pt. The lines represent the average value.

## APPENDIX C: ANGULAR DEPENDENT ST-FMR

The lineshape analysis which uses the ratio of symmetric (S) to antisymmetric (A) component at one fixed angle of $\phi = 45°$, could be a hindrance in revealing the complete picture of exerted spin-orbit torques [21, 36]. There may be hidden effects apart from SHE, such as effective field with different spin polarization, poor device designs, Nernst heating, etc. which may serve as an artifact and lead to an unreliable assessment of spin-orbit torque. So, we performed the angular ST-FMR measurements by varying the angle between applied $H_{ext}$ and length of device axis from $\phi = 0^o$ to $360^o$. By deconvolution of $V_{mix}$ using Eq. (1) into S and A , we fitted the data with the anticipated $\sin2\phi\cos\phi$ for the implanted sample Pt(N) $2\times10^{16}$, Pt(N) $5\times10^{16}$ and Pt(N) $1\times10^{17}$ (Fig. 7(a), (b), (c)). It shows that the SOT traces similar lineshape when the magnetization is rotated by $180^o$ which implies that there is no breaking of the twofold ($180^o+\phi$) and mirror ($180^o-\phi$) symmetries of torques, affirming SHE as the only origin of the rectified voltage [21] obtained by ST-FMR. The $\cos\phi$ arises from $\tau_{DL}$ and $\tau_{OFT}$ in S and A, respectively, while $\sin2\phi$ arises from AMR.

## APPENDIX D: SPIN PUMPING CONTRIBUTION

Spin pumping contribution $V_{sp}$ in the symmetric component of the ST-FMR spectrum may have a role to play in the high values of $\alpha$, $g_{eff}^{\uparrow\downarrow}$ and $\theta_{DL}$ for Pt(N) 5×10$^{16}$, and therefore it might be naive to not identify this contribution. To confirm that the enhanced $\alpha$, $g_{eff}^{\uparrow\downarrow}$ and $\theta_{DL}$ in N-implanted sample may be attributed to the enhanced DLT, and is not due to the contribution of spin pumping voltage in the symmetric component of ST-FMR spectrum, we investigated the spin pumping contribution $V_{sp}$ using the derived values of $g_{eff}^{\uparrow\downarrow}$ and $\theta_{DL}^{LS}$ by [11]:

$$V_{SP} = \frac{\theta_{DL} l \lambda_{sd}}{d\sigma_{Pt(N)} + t\sigma_{NiFe}} \tanh\left(\frac{d}{2\lambda_{sd}}\right)\left(\frac{2e}{h}\right) j_S \sin(\phi)\,, \qquad (9)$$

where, l is the length of the device, $\lambda_{sd}$ is the spin diffusion length of Pt(N) layer, d and $\sigma_{Pt(N)}$ are the thickness and conductivity of Pt(N), t and $\sigma_{NiFe}$ are the thickness and conductivity of NiFe, and $\phi$ is the angle between $H_{ext}$ and $I_{rf}$ applied in the longitudinal direction of device (45°), $j_S$ is the spin current density from the precessing NiFe into the Pt(N), given as $j_S = \frac{h}{2} f \sin^2(\theta_C) g_{eff}^{\uparrow\downarrow}$, where the precession cone angle $\theta_C$ is given by $\theta_C = \frac{1}{dR/d\phi}\frac{2}{I_{rf}}\sqrt{S^2 + A^2}$, $dR/d\phi$ is obtained from anisotropic magnetoresistance, and $I_{rf}$ is the current in the device. The ratio of the spin pumping voltage $V_{sp}$ to the symmetric component S is found to be 0.39% for Pt(N) 2×10$^{16}$, 0.49% for Pt(N) 5×10$^{16}$ and 0.25% for Pt(N) 1×10$^{17}$, which are all less than 1%. Figure 7(d) shows S and $V_{sp}$ plotted for Pt(N) 5×10$^{16}$ as a function of frequency, confirming the negligible contribution of $V_{sp}$ as compared to S.

## APPENDIX E: SPIN MIXING CONDUCTANCE

Spin mixing conductance $g_{eff}^{\uparrow\downarrow}$ is an important parameter that provides a better picture of transversely generated $j_S$ created at the FM/HM interface. Based on the theory of spin pumping, assuming that there is no significant spin memory loss, the $g_{eff}^{\uparrow\downarrow}$ [21] can be estimated from the linewidth difference δ of ST-FMR spectra ($\delta = \Delta H_{Pt(N)/NiFe} - \Delta H_{NiFe}$), and is given by:

$$g_{eff}^{\uparrow\downarrow} = \left(\frac{\gamma}{2\pi f}\right)\left(\frac{4\pi M_s t\delta}{g\mu_o\mu_B}\right), \tag{8}$$

where, g is the Landé g factor, $\mu_o$ is the permeability of free space, $M_S$ is the saturation magnetization of NiFe and $\mu_B$ is the Bohr magneton constant. The average value of $g_{eff}^{\uparrow\downarrow}$ is found to be $2.08 \times 10^{19}$ $m^{-2}$ for Pt(N) $2\times10^{16}$, $3.10 \times 10^{19}$ $m^{-2}$ for Pt(N) $5\times10^{16}$ and $2.57 \times 10^{19}$ $m^{-2}$ for Pt(N) $1\times10^{17}$ as shown by the dashed lines in Fig. 7 (e). Most importantly, Pt(N) $5\times10^{16}$ was found to have a higher $g_{eff}^{\uparrow\downarrow}$ value in comparison to Pt(N) $2\times10^{16}$ and Pt(N) $1\times10^{17}$. For a HM layer much thicker than its spin diffusion length, $\theta_{DL}$ is found to be proportional to $g_{eff}^{\uparrow\downarrow}$ [37]. This is linked to the fact that the Gilbert damping parameter $g_{eff}^{\uparrow\downarrow}$ and $\theta_{DL}$ of Pt(N) $5\times10^{16}$ is found to be larger than that of Pt(N) $2\times10^{16}$ and Pt(N) $1\times10^{17}$.

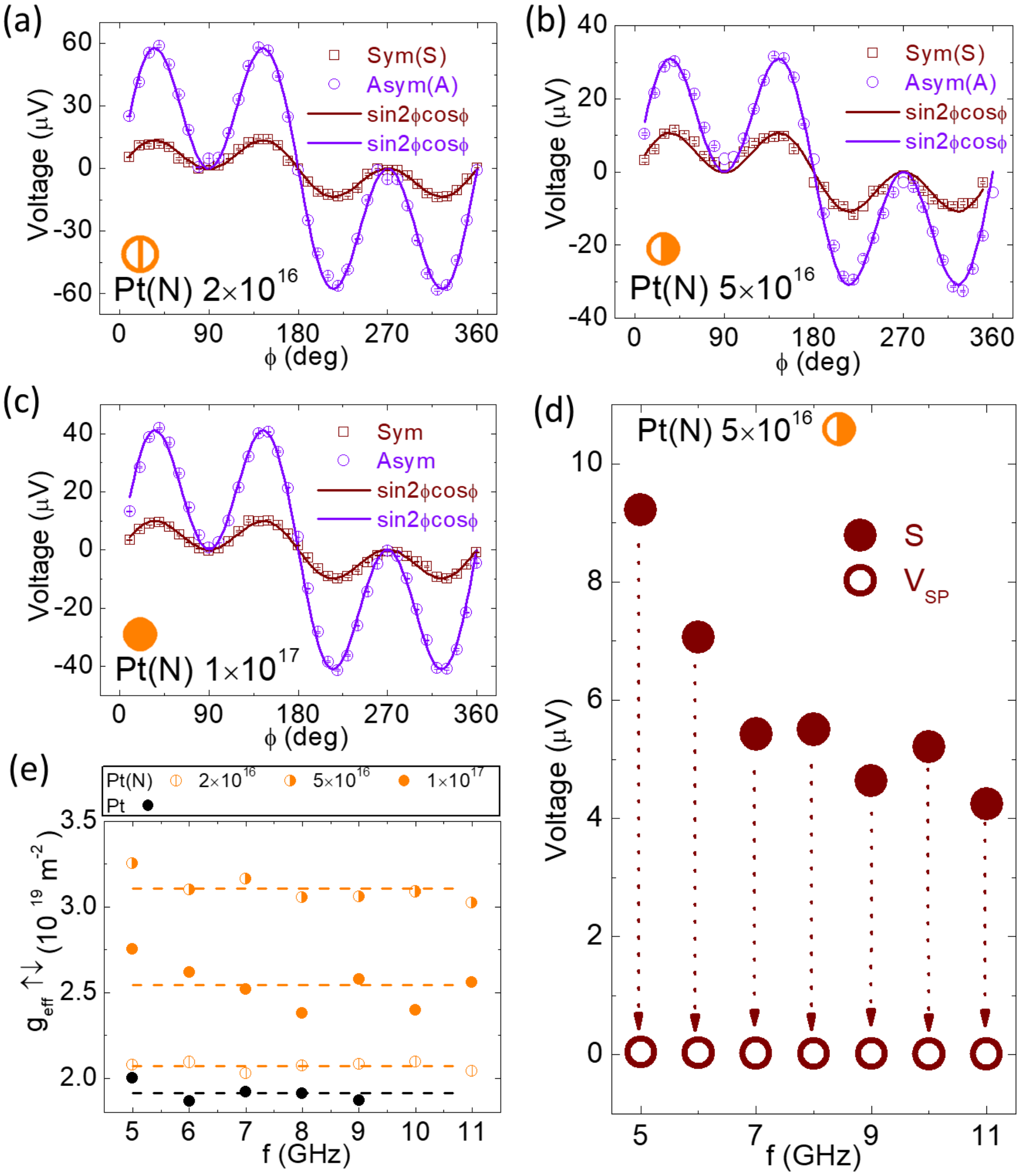

Fig. 7: (a), (b), (c) Angular dependence of symmetric (S) and antisymmetric (A) components in ST-FMR spectra for Pt(N) 2×10$^{16}$, Pt(N) 5×10$^{16}$, and Pt(N) 1×10$^{17}$ (f = 5 GHz) with solid lines fitted by sin 2ϕ cos ϕ. (d) Symmetric component (S) and spin pumping contribution ($V_{sp}$) plotted as a function of frequency for Pt(N) 5×10$^{16}$. (e) $g_{eff}^{\uparrow\downarrow}$ obtained as a function of f for Pt(N) 2×10$^{16}$, Pt(N) 5×10$^{16}$, Pt(N) 1×10$^{17}$ along with pure Pt. The dashed lines represent the average value.